\documentclass[preprint,showpacs,showkeys,preprintnumbers,amsmath,amssymb]{revtex4}
\usepackage{natbib}
\usepackage{graphicx}% Include figure files
\usepackage{dcolumn}% Align table columns on decimal point
\usepackage{bm}% bold math
\usepackage{longtable}% long table

\begin{document}

%\preprint{APS/123-QED}

\title{A Density Functional Study of Bare and Hydrogenated Platinum Clusters}

\author{Ali Sebetci}
 \email{asebetci@cankaya.edu.tr}
\affiliation{Department of Computer Engineering, \c{C}ankaya
University, 06530 Balgat Ankara, Turkey}

\date{\today}% It is always \today, today,

\begin{abstract}
We perform density functional theory calculations using Gaussian
atomic-orbital methods within the generalized gradient approximation
for the exchange and correlation to study the interactions in the
bare and hydrogenated platinum clusters. The minimum-energy
structures, binding energies, relative stabilities, vibrational
frequencies and the highest occupied and lowest unoccupied
molecular-orbital gaps of Pt$_{n}$H$_{m}$ ($n$=1-5, $m$=0-2)
clusters are calculated and compared with previously studied pure
platinum and hydrogenated platinum clusters. We investigate any
magic behavior in hydrogenated platinum clusters and find that
Pt$_{4}$H$_{2}$ is more stable than its neighboring sizes. Our
results do not agree with a previous conclusion that 3D geometries
of Pt tetramer and pentamer are unfavored. On the contrary, the
lowest energy structure of Pt$_{4}$ is found to be a distorted
tetrahedron and that of Pt$_{5}$ is found to be a bridge site capped
tetrahedron which is a new global minimum for Pt$_{5}$ cluster. The
successive addition of H atoms to Pt$_{n}$ clusters leads to an
oscillatory change in the magnetic moment of Pt$_{3}$ - Pt$_{5}$
clusters.
\end{abstract}

%\keywords{}

%Use showkeys class option if keyword
\maketitle

\section{INTRODUCTION}

The study of hydrogen interaction with metals has gained an
increased interest in the last
decade~\cite{Okamoto,Poulain,Bartczak,Khanna,Musaev,Mainardi,Kumar}.
The use of metal and semiconductor clusters as components of
nanodevices, the development of cluster-based materials and the
catalytic properties of clusters are the main reasons for the
studies of interaction of atoms and molecules with clusters.
Understanding hydrogen interaction with clusters is important
because of two basic reasons: first, there is a great interest in
developing novel hydrogen absorbing nanomaterials for fuel cell
applications, second, many of the organic materials and biological
systems contain hydrogen which are involved important catalytic
reactions. In particular, platinum is one of the most important
ingredients in the heterogeneous catalysis of hydrogenation as well
as in the catalysis of the CO, NO$_{x}$, and hydrocarbons. Moreover,
detailed investigations of the interaction between platinum clusters
with hydrogen can contribute to the design of useful hydrogen
storage devices. Therefore, it is of interest to understand the
interaction of one and two H atoms on platinum clusters.

A brief summary of the previous theoretical and experimental studies
on the pure platinum clusters can be found in our recent
publications~\cite{Ali}. The following references can be added to
them: Gr\"{o}nbeck and Andreoni~\cite{Gronbeck} have presented a
density functional theory (DFT) study of Au$_{2}$-Au$_{5}$ and
Pt$_{2}$-Pt$_{5}$ clusters in the neutral and anionic states. Xiao
and Wang~\cite{Xiao} have studied Pt clusters of up to 55 atoms
using DFT with a plane wave basis set and Tian et al.~\cite{Tian}
have reported the geometrical and electronic structure of the
Pt$_{7}$ cluster in a DFT study with a Gaussian type basis set. One
of the earliest {\it ab initio} studies on the Pt-H$_{2}$ reaction
was published by the group of Poulain~\cite{Poulain2} in 1986. The
same group carried out theoretical studies on the
Pt$_{2}$-H$_{2}$~\cite{Poulain3}, and
Pt$_{4}$-H$_{2}$~\cite{Poulain} interactions later. Nakasuji et
al.~\cite{Nakasuji} accomplished an {\it ab initio} study on the
reactions of the hydrogen molecule with small Pt$_{n}$ ($n$=1-3)
clusters. In 1987, Balasubramanian's~\cite{Bala1} calculations on
electronic states and potential energy surfaces of PtH$_{2}$ has
been published. Later, he and Feng~\cite{Bala2} have reported a
study on the potential energy surfaces for the Pt$_{2}$-H and Pt-H
interactions. The Pt$_{3}$-H$_{2}$ interactions have been studied by
Dai et al.~\cite{Dai} and Cui et al.~\cite{Cui}.

On the experimental side, molecular beam techniques have been used
to investigate the hydrogen molecule interaction with metal
surfaces~\cite{Samson}. The chemical reactivity of metal clusters in
gas phase with hydrogen and deuterium has been studied with the
laser vaporization technique~\cite{Kaldor}. In a most recent study,
Andrews et al.~\cite{Andrews} have investigated the reaction of Pt
atoms with H$_{2}$ in a laser-ablation experiment and have also
performed some DFT calculations. In the experiments it is observed
that the reactivity of the small metal clusters with the hydrogen
and deuterium molecules strongly depends on the size of the small
metal clusters~\cite{Knick}. Luntz et al.~\cite{Luntz} found that
there are reaction paths with or without very low barriers, leading
to the H$_{2}$ dissociation on the Pt (111) surfaces.

In the present work, we report the results of first-principles
total-energy calculations on small platinum clusters with up to two
H atoms. We perform DFT calculations using Gaussian atomic-orbital
methods within the generalized gradient approximation for the
exchange and correlation to study the interactions in the bare and
hydrogenated platinum clusters. The possible minima and ground
states, binding energies (BE), relative stabilities, and the highest
occupied and the lowest unoccupied molecular-orbital (HOMO-LUMO)
gaps of the Pt$_{n}$H$_{m}$ ($n$=1-5, $m$=0-2) clusters have been
calculated. Vibrational frequency calculations for each optimized
structure has been carried out to differentiate local minima from
transition states. The results are compared with previously studied
pure platinum and hydrogenated platinum clusters. Systematic growth
in the atom number allows one to monitor the changes in the
structure, dynamics, and electronic properties involved with the
size of the clusters. This can provide an understanding of the
nature of interaction between hydrogen and Pt clusters.

\section{COMPUTATIONAL METHOD}

DFT calculations have been performed by using version 4.7 of the
NWChem program~\cite{NWChem}. We have employed the
LANL2DZ~\cite{LAN} basis set for the Pt atoms, which uses
relativistic effective core potentials (ECP) to reduce the number of
electrons explicitly considered in the calculation, and 6-311++G**
basis set for the H atoms. The ECP parameters for platinum and the
basis set, as well as similar basis sets to the one employed in this
study for the hydrogen have been successfully used in many previous
studies~\cite{Kumar,Tian,Poulain2,Poulain3,Andrews}. To examine the
effect of the choice of the exchange-correlation functional, we have
calculated the properties of H and Pt dimers and those of PtH
diatomic molecule by using different functionals: We have employed
PW91PW91~\cite{Perdew} and BPW91~\cite{Becke1,Perdew} functionals
within the generalized gradient approximation (GGA),
SVWN5~\cite{Slater} functional within the local density
approximation (LDA), and B3LYP~\cite{Becke2,Perdew} and
B98PW91~\cite{Becke3,Perdew} methods as hybrid functionals. At the
end of the examination which is discussed in the next section, we
have chosen the BPW91 method to study Pt$_{n}$H$_{m}$ clusters. The
default NWChem geometry optimization convergence criteria were used
in all cases. Initial structures have been relaxed without imposing
any symmetry constraints. The final structures therefore include
possible Jahn-Teller distortions. Whenever the optimized structure
has had a symmetry other than $C_{1}$, the relaxation process has
been repeated with that symmetry constraints starting from the
optimized geometry in order to get the electronic state.
Spin-polarized calculations have been done for the first two
multiplicities of the clusters up to Pt$_{4}$H. Thus, singlet and
triplet states have been worked out for those having even number of
electrons and doublet and quartet states have been worked out for
those having odd number of electrons. In addition, we have also
studied the quintet states of Pt$_{4}$H$_{2}$, Pt$_{5}$ and
Pt$_{5}$H$_{2}$ clusters, and sextet states of Pt$_{5}$H clusters.
The reason for this is discussed in Sec.~\ref{Moment}. All figures
were produced by the ChemCraft graphics program~\cite{ChemCraft}.

\subsection{Effect of Exchange-Correlation Functional}

Table~\ref{table1} displays the calculated and experimental values
of the bond lengths, vibrational frequencies and BEs for H and Pt
dimers and for PtH diatomic molecule. The experimental results of
H$_2$, Pt$_2$, and PtH are from Refs.~\cite{Huber1, Huber2},
Refs.~\cite{Airola,Fabbi,Taylor}, and Refs.~\cite{Huber1,McCarthy},
respectively. In our calculations, the hybrid method B3LYP has
produced the nearest bond lengths to the experimental values of both
H$_2$ and PtH molecules. However, it has resulted in the worst bond
length (2.47 $\AA$) for Pt$_2$ (experimental value is 2.33 $\AA$).
Therefore, it highly underestimates the BE of Pt dimer by
calculating only a half of the experimental value. When a second
hybrid functional B98PW91 is considered, it can be seen in
Table~\ref{table1} that although the results for Pt$_2$ are improved
as compared to the results of the B3LYP method, it still
overestimates the bond length and underestimates the BE of the dimer
significantly. On the other hand, the SVWN5 LDA method has produced
very good results for the bond lengths of all of the three
molecules. However, when the BEs are taken into account, it is not
favorable, too. It highly overestimates Pt$_2$ and PtH BEs. We have
tried two GGA methods for the properties of the three diatomic
molecules: PW91PW91 and BPW91. Both of them have resulted in nearly
the same bond lengths which are excellent for H$_2$ and PtH, and
very good for Pt$_2$. The errors in the bond length of Pt$_2$
produced by GGA methods are only 1.7\%. They have also calculated
reasonably good BEs for Pt$_2$ and PtH. PW91PW91 functional has
produced a slightly better result for the BE of PtH than BPW91.
However, the later has calculated much better Pt$_2$ BE than the
former. Therefore, we have decided to employ BPW91 functional in the
rest of our calculations. The error produced by the chosen method in
the BE of H molecule (9.5\% ) is not very important in the present
study since the H atoms in the Pt$_{n}$H$_{m}$ clusters do not bind
together as we discuss in the following sections. As a result, we
have performed all the remaining computations of the hydrogenated Pt
clusters at the BPW91/LANL2DZ and BPW91/6-311++G** levels of theory.

\section{RESULTS AND DISCUSSION}

\subsection{Bare Pt Clusters: Pt$_{2}$-Pt$_{5}$}

We start to report the obtained results with bare Pt clusters.

Pt$_{2}$: It can be seen from Table~\ref{table1} and
Table~\ref{table2} that the chosen DFT method produces reasonably
good results for the BE, vibrational frequency and the bond length
of the Pt dimer. We have determined the electronic state of the
dimer as $^{3}\Delta_{g}$ which is not consistent with
Balasubramanian's~\cite{Bala1} result that a $^{3}\Sigma_{g}^{-}$ is
the ground state. The singlet state turns out to be 1.39 eV less
stable than the triplet. Both of the references~\cite{Gronbeck} and
\cite{Xiao} have reported the ground state of the dimer as a
triplet, however none of them has given any electronic state. Xiao
and Wang~\cite{Xiao} have calculated 3.52 eV BE, 2.34 $\AA$ bond
distance and 0.81 eV HOMO-LUMO gap in their DFT study with ultrasoft
pseudopotentials. The best energy and bond length values of
Gr\"{o}nbeck and Andreoni~\cite{Gronbeck} obtained by employing BLYP
exchange correlation functional are 3.58 eV and 2.32 $\AA$,
respectively. For both of these studies, the bond distance values
are closer to the experimental data (2.33 $\AA$) than our result
(2.37 $\AA$), however the method employed in this work results
better in the energy since it calculates 3.36 eV BE which is closer
to the experimental data (3.14 eV) than the previous calculations.
Balasubramanian~\cite{Bala1} has reported the obtained Pt-Pt
distance as 2.46 $\AA$ which is the worst among all of the above
mentioned references. We have calculated 1.32 eV HOMO-LUMO gap which
is significantly higher than the one given in Ref.~\cite{Xiao}. Our
result for the $\beta$ spin HOMO-LUMO gap (0.33 eV) is in agreement
with the calculation of Tian et al.~\cite{Tian} (0.3 eV).

\begin{figure}
\includegraphics[bb= 0 0 50.02cm 21.27cm, scale=0.25]{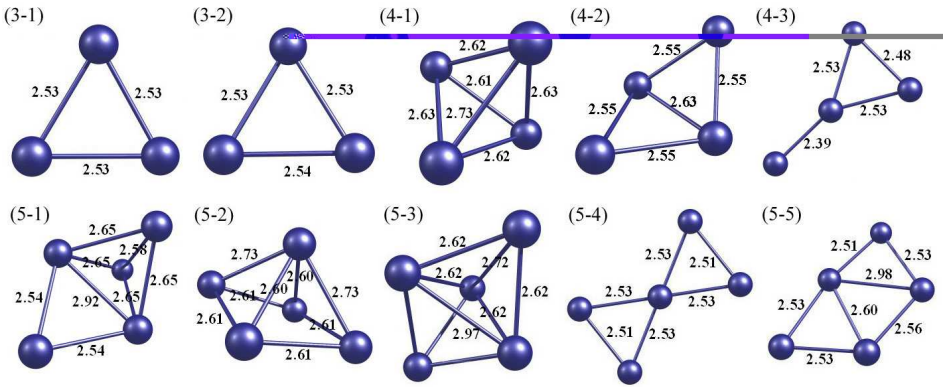}
\caption{\label{Pt}Relaxed structures of Pt$_n$ ($n$=3-5). All
distances are in $\AA$. }
\end{figure}

Pt$_{3}$: The isomeric structures, symmetries, ground electronic
states, total BEs, HOMO-LUMO gaps ($\alpha$ spin), and the highest
and the lowest vibrational frequencies of Pt$_{2}$-Pt$_{5}$ clusters
are given in Table~\ref{table2}. The pictures of these isomers and
the bond lengths (in $\AA$) in these structures are presented in
Fig.~\ref{Pt}. We have studied the equilateral and isosceles
triangle structures of the Pt trimer, and found out that the
optimized bond lengths of these two geometries are very similar (see
Fig.~\ref{Pt} (3-1) and (3-2)). Therefore, the energy separation
between them is very small. All Pt-Pt distances in the equilateral
triangle are 2.53 $\AA$ which can be compared with the result of
Gr\"{o}nbeck and Andreoni~\cite{Gronbeck} (2.41 $\AA$). The ground
electronic state of the $D_{3h}$ symmetry is $^{3}E''$ whereas that
of $C_{2v}$ symmetry is $^{3}B_{2}$. The BEs and the HOMO-LUMO gaps
for both of these structures are 6.57 eV and 1.05 eV, respectively.

Pt$_{4}$: We have identified three stable structures of Pt$_{4}$: a
distorted tetrahedron (4-1), an out of plane rhombus (4-2), and a
planar Y-like shape (4-3). A square isomer has been found to be a
first order transition state in our calculations. The distorted
tetrahedron with $C_{2}$ symmetry and $^{3}B$ electronic state is
0.1 eV more stable than the rhombus, whereas the BE of the rhombus
is 0.29 eV higher than the BE of the third isomer. The HOMO-LUMO
gaps and the lowest and highest vibrational frequencies of these
clusters can be found in Table~\ref{table2}. Previously,
Gr\"{o}nbeck and Andreoni~\cite{Gronbeck} have reported the global
minimum to be a slightly out of plane rhombus in the triplet. Xiao
and Wang~\cite{Xiao} have found a distorted tetrahedron with $C_{s}$
symmetry as the lowest energy structure in their plane wave
calculations. Dai and Balasubramanian~\cite{Dai2} have performed
MRSDCI calculations and a regular tetrahedron in a triplet state was
determined to be the ground state.

Pt$_{5}$: A bridge site capped tetrahedron (5-1), a rhombus pyramid
(5-2), a trigonal bipyramid (5-3), an X-like (5-4) and an W-like
(5-5) geometries are the low-energy structures of Pt$_{5}$. This is
the first time, to the best of our knowledge, that the bridge site
capped tetrahedron (5-1) is identified as the global minimum of
Pt$_{5}$. It is in the quintet state and 0.09 eV more stable than
the second isomer. The second isomer (rhombus pyramid) is also in
the quintet state, whereas the other three locally stable structures
are in triplet. The HOMO-LUMO gap (0.69 eV) of the lowest energy
structure (5-1) is also higher than that of the other isomers.
Therefore, it is expected to be more stable than the other
structures. The BE of the rhombus pyramid is 0.07 eV higher than the
BE of the trigonal bipyramid. The energy separation of the almost
planar X-like and W-like structures from the 3D configurations is
more than 0.04 eV. In Ref.~\cite{Gronbeck}, BLYP calculations have
predicted the lowest energy isomer to be the planar W-like shape and
LDA calculations have placed a distorted square pyramid at low
energies. In Ref.~\cite{Xiao}, the trigonal bipyramid has been found
as the global minimum in the quintet state.

According to the results presented here, the claim that 3D
geometries for Pt tetramer and pentamer are unfavored is not true.
In addition, the HOMO-LUMO gaps for Pt$_{2}$-Pt$_{5}$ obtained in
our calculations (1.32, 1.05, 0.44, and 0.69 eV, respectively) are
generally much greater than the previous calculations (0.7, 0.2,
0.3, and 0,2 eV in Ref.~\cite{Gronbeck}, and 0.81, 0.034, 0.28, and
0.63 eV in Ref.~\cite{Xiao}).

\subsection{One and Two Hydrogens on Pt Clusters}

\begin{figure}
\includegraphics[bb= 0 0 64cm 84cm, scale=0.25]{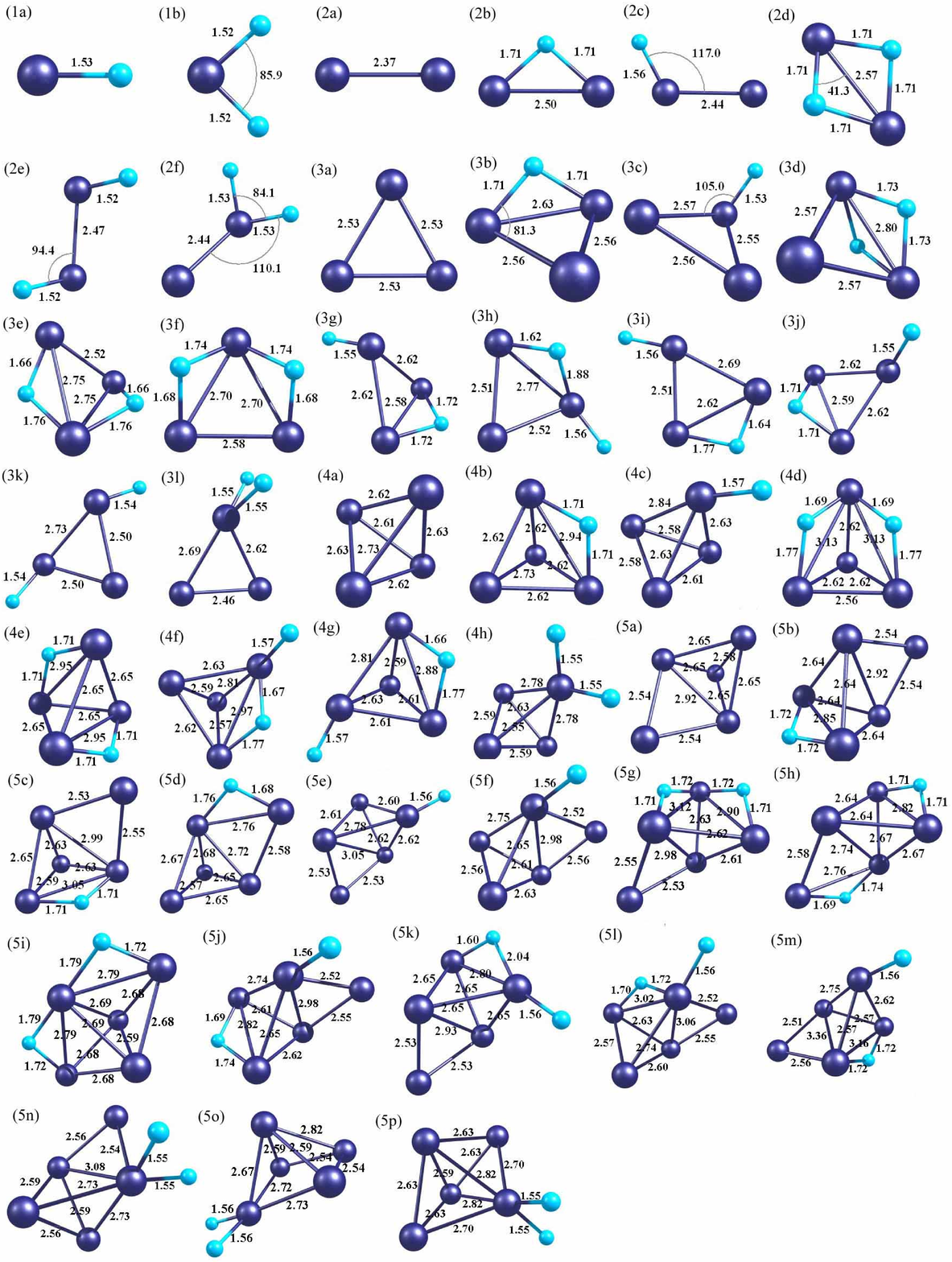}
\caption{\label{Pt-H}Relaxed structures of Pt$_n$H$_m$ ($n$=1-5 and
$m$=0-2). Distances are in $\AA$ and angles are in degree.}
\end{figure}

The relaxed structures and bond lengths of one and two H atoms on
small Pt clusters are shown in Fig.~\ref{Pt-H}, while the
symmetries, electronic states, BEs, HOMO-LUMO gaps ($\alpha$ spin)
and the highest and the lowest vibrational frequencies are given in
Table~\ref{table3}.

PtH: Considering first the interaction of H with a Pt atom will be
instructive. We have calculated the BE of the PtH molecule in the
doublet state as 3.28 eV, which is close to the experimental
data~\cite{Huber1} of 3.44 eV. In fact, as discussed by
Balasubramanian and Feng~\cite{Bala2} the experimental value should
be regarded as only an upper bound. They have reported the ground
state BE as 3.11 eV by employing the CASSCF/SOCI/RCI method in
1989~\cite{Bala2}. We have identified the electronic ground state of
the molecule as $^{2}\Delta$ which is in agrement with both the
experimental~\cite{McCarthy} and Balasubramanian's theoretical
results~\cite{Bala2}. Andrews et al.~\cite{Andrews} have
investigated PtH$_{n}$ ($n$=1-3) clusters by employing the B3LYP and
BPW91 density functionals with 6-311+G**, 6-311++G(3df,3pd) and
aug-cc-pVTZ basis sets for H, and LANL2DZ and SDD pseudopotentials
for Pt. In their study, the ground state has been determined as
$^{2}\Sigma^{+}$, however no BE has been given. We have computed the
bond length of PtH as 1.528 $\AA$ which is also in very good
agreement with the experimental value~\cite{McCarthy} of 1.529
$\AA$. It is determined as 1.55 $\AA$ in Ref.~\cite{Bala2} and as
1.531 $\AA$ in Ref.~\cite{Andrews}. When the vibrational frequency
is considered, the experimental data~\cite{McCarthy} is 2294
cm$^{-1}$. Our result is 2420 cm$^{-1}$ whereas Ref.~\cite{Bala2}
and Ref.~\cite{Andrews} have reported 2177 and 2370 cm$^{-1}$,
respectively. It should be noted here that all frequencies given in
this work are determined without using any scale factor. If they
were scaled by a proper factor for the chosen method, they would be
more accurate. The HOMO-LUMO gap of PtH is 3.02 eV which is the
largest gap among all the clusters we have studied here. Since the
total energy of Pt$_{2}$ and H$_{2}$ (7.66 eV) is bigger than the
energy of 2PtH (6.56 eV), one can expect the abundance of the PtH
molecule in an environment of atomic hydrogen rather than in an
environment of molecular hydrogen.

PtH$_{2}$: PtH$_{2}$ has a waterlike structure as shown in
Fig.~\ref{Pt-H} (1b) with an angle of 85.9$^\circ$. This angle was
reported previously as 85$^\circ$ by Balasubramanian~\cite{Bala1}
and as 85.7$^\circ$ by Andrews et al.~\cite{Andrews}. As in the case
of the angle, our result for the bond length (1.517 $\AA$) is in
very good agreement with the previous calculation (1.520
$\AA$)~\cite{Andrews} . All of these studies (Ref.~\cite{Bala1},
Ref.~\cite{Andrews}, and the present paper) identify the ground
state of the PtH$_{2}$ molecule to be $^{1}A_{1}$. It can be seen
from Table~\ref{table3} that the BE of H on PtH (3.58 eV) is 0.30 eV
higher than the BE of H on Pt atom (3.28 eV). Thus, PtH$_{2}$ is a
bit more stable than PtH when H separations from these molecules are
considered. Although the HOMO-LUMO gap of PtH$_{2}$ (2.49 eV) is
smaller than that of PtH, it is still a big gap. We have found
PtH$_{2}$ to be 6.86 eV more stable than the ground state Pt atom
and two ground state H atoms. In order to dissociate H$_{2}$
molecule from PtH$_{2}$, 2.55 eV is needed. This energy may be
compared with earlier calculations: 2.15 eV~\cite{Andrews}, 1.73
eV~\cite{Nakasuji}, and 2.04 - 2.21 eV~\cite{Bala1}. We have
determined the lowest, middle and the highest vibrational
frequencies as 793, 2455, 2490 cm$^{-1}$. In Ref.~\cite{Andrews},
the corresponding frequencies are computed in the following ranges
by employing different methods and basis sets: 784 - 805, 2414 -
2456, 2455 - 2513 cm$^{-1}$. The H-H distance in PtH$_{2}$ is 2.07
$\AA$, which is significantly greater than the bond length of
H$_{2}$ (0.74 $\AA$). Therefore, the interaction between Pt and
H$_{2}$ is dissociative.

Pt$_{2}$H: There are two possible ways of H bonding on the Pt dimer:
on the bridge site (Fig.~\ref{Pt-H} (2b) which is an isosceles
triangle) and on the top site (Fig.~\ref{Pt-H} (2c) which is a
planar V-like shape with $C_{s}$ symmetry). Different electronic
states of the bridge structure have been studied in
Ref.~\cite{Bala2} before and a $^{2}A_{2}$ state has been identified
as the ground state, which is in full agreement with our result.
2.53 $\AA$ of the Pt-Pt distance and 1.75 $\AA$ of the Pt-H distance
given in that study can be compared with the present results of 2.50
$\AA$ and 1.71 $\AA$, respectively. We have determined the ground
state of the top structure to be a $^{4}A''$ state. The energy
difference between the quartet and doublet states of this geometry
is 0.50 eV. The BEs of H on the bridge and top sites of Pt$_{2}$ are
2.73 eV and 2.25 eV, respectively. Since these energies are
particularly high it can be concluded that Pt$_{2}$ is a reactive
molecule. The bonding on the bridge site is 0.48 eV more favorable
than the bonding on the top site. On the other hand, the HOMO-LUMO
gap of (2c) structure (2.57 eV) is much greater than that of (2b)
structure (1.26 eV). Thus, the further reaction of the isosceles
triangle to make Pt$_{2}$H$_{2}$ can occur easier than that of the
V-like structure. In the V-like structure, the Pt-Pt and Pt-H
distances are reduced to 2.44 $\AA$ and 1.56 $\AA$.

Pt$_{2}$H$_{2}$: We have identified three stable structures of
Pt$_{2}$H$_{2}$; (i) both H atoms are on the bridge site
(Fig.~\ref{Pt-H} (2d)), (ii) both of them are on the top sites of
different Pt atoms (Fig.~\ref{Pt-H} (2e)), and (iii) both of them
are on the top site of the same Pt atom (Fig.~\ref{Pt-H} (2f)). An
initial geometry of a bridge-top configuration takes the (2e) form
at the end of the optimization process. The singlet states of all of
these three structures are more stable than their triplet states.
Among the three local minima, (2e) form has the biggest BE (9.29
eV). The HOMO-LUMO gap of the (2f) form (0.09 eV) is significantly
small, therefore it is expected to react further to make
Pt$_{3}$H$_{2}$ or Pt$_{2}$H$_{3}$. The H-H distance in (2f) is 2.05
$\AA$ which is again much bigger than the bond distance in the H
molecule. Therefore, the H adsorption on Pt dimer is also
dissociative.

Pt$_{3}$H: Interaction of H on Pt$_{3}$ is slightly favorable on the
bridge site (Fig.~\ref{Pt-H} (3b)) as compared to the top site
(Fig.~\ref{Pt-H} (3c)) by 0.06 eV. The HOMO-LUMO gap in the (3b)
form (0.66 eV) is also greater than the gap in the (3c) form (0.19
eV). The ground states of both of these structures are doublet. The
H BEs in these geometries are 2.63 eV and 2.57 eV respectively,
which are relatively small when compared to the most of the BEs in
the other Pt$_{n}$H$_{m}$ clusters. The Pt-H distance in the bridge
site (1.71 $\AA$, (3b)) is very similar to the corresponding bridge
site distances of the previous clusters, that is Pt-H distances in
Pt$_{2}$H (2b) and Pt$_{2}$H$_{2}$ (2d). The same is true for the
typical 1.53 $\AA$ Pt-H top site distance in the (3c) form which is
very close to the Pt-H distances in the (1a), (1b), (2e), and (2f)
structures. The only exception of this is the 1.56 $\AA$ Pt-H bond
length in the (2c) form which can be related to the fact that (2c)
structure is a quartet although all the other Pt$_{n}$H clusters
listed above are doublets. In our calculations, we have not found
any local minima of Pt$_{3}$H in which H has a coordination number
of three or more.

Pt$_{3}$H$_{2}$: We have identified 9 different local minima of
Pt$_{3}$H$_{2}$; three bridge-bridge configurations (Fig.~\ref{Pt-H}
(3d)-(3f)), four bridge-top configurations (Fig.~\ref{Pt-H}
(3g)-(3j)), and two top-top configurations (Fig.~\ref{Pt-H} (3k),
(3l)). In the (3d) geometry the two H atoms are on the same bridge
site, whereas in the (3e) and (3f) forms they are on the neighboring
sites. The difference between (3e) and (3f) is that in the former
case H atoms are on the opposite sides of the Pt$_{3}$ plane,
whereas in the later case they are not. Among the four bridge-top
configurations, (3h) is the only one having a Pt atom bonding to
both of the H atoms. In the (3g) and (3j) geometries the bridge and
top site H atoms are on the opposite and on the same sides of the
Pt$_{3}$ plane, respectively. In the (3i) case, all the atoms of the
cluster are on the same plane. In one of the top-top configuration,
the H atoms are bonded to different Pt atoms (3k). They are bonded
to the same Pt atom in the other one (3l). The top-top
configurations have greater BEs than all the other isomers (see
Table~\ref{table3}). When the two H atoms are bonded to the top site
of the same Pt atom (3l), the ground state is a triplet which is
different from the singlet (2f) structure of Pt$_{2}$H$_{2}$. The
(3l) structure has the greatest BE and HOMO-LUMO gap, and therefore
it is the most stable Pt$_{3}$H$_{2}$ cluster that we have
investigated in the present work. The H-H distance in it is 1.75
$\AA$. The BE for 2H is 5.51 eV and it is significantly higher than
the dissociation energy of H$_{2}$ (4.75 eV). Accordingly, hydrogen
is likely to be dissociated on Pt$_{3}$ as well as on Pt$_{2}$.
Among the bridge-top configurations, (3h) structure has higher BE
than the others. The bridge-top isomers of the Pt$_{3}$H$_{2}$
cluster are generally more stable than its bridge-bridge isomers
except that (3f) geometry has greater BE than (3i).

Pt$_{4}$H \& Pt$_{4}$H$_{2}$: H prefers to bind to the ground state
Pt tetramer on the top site. The isomer with H on the top site
(Fig.~\ref{Pt-H} (4c)) is a quartet and its BE is 0.3 eV greater
than that of the isomer with H on the bridge site (Fig.~\ref{Pt-H}
(4b)) which is a doublet. The HOMO-LUMO gap of (4c) is also greater
than that of (4b) (see Table~\ref{table3}). Thus, (4c) structure is
expected to be more stable. In the case of 2H, the most stable
structure is a top-top configuration (Fig.~\ref{Pt-H} (4h)) which is
similar to the Pt$_{3}$H$_{2}$ case. The HOMO-LUMO gap of this
structure is 0.94 eV and it is higher than that of all the other
Pt$_{4}$H$_{2}$ isomers. The H-H distance in (4h) is 1.95 $\AA$. The
2H BE is 5.80 eV. Thus, the H absorbtion is again dissociative. The
BEs of the neighbor bridge-bridge configuration (4d) and the
opposite bridge-bridge configuration (4e) are nearly the same.
However, (4d) is a triplet whereas (4e) is a singlet. For the
bridge-top configurations, the neighboring structure (4f) is 0.1 eV
more stable than the opposite structure (4g).

Pt$_{5}$H \& Pt$_{5}$H$_{2}$: H can bind to the ground state Pt
pentamer on three different bridge and two different top sites. Top
site bonding is favorable in BE: (5f) is the lowest energy
structure. However the HOMO-LUMO gaps of the bridge site bonded
structures are higher than those of top site bonded structures. All
Pt$_{5}$H isomers in Table~\ref{table3} are in the quartet states.
For the bonding of 2H, we have identified three bridge-bridge, four
bridge-top, and three top-top locally stable configurations. In each
of these 10 optimization processes, we have started from different H
locations on the lowest energy structure (5-1) of the Pt$_{5}$
cluster. However, in three of them (5-1) configuration of the 5 Pt
atoms changed into some other topologies at the end of the
relaxation: The (5i) and (5p) geometries are based on trigonal
bipyramid and the (5o) resembles to rhombus pyramid. In agreement
with the general trend, the most stable Pt$_{5}$H$_{2}$ isomer is
the (5p) top-top structure with the highest BE and the highest
HOMO-LUMO gap. The H-H distance in (5p) is 1.91 $\AA$ and the 2H BE
is 5.59 eV. Thus, the H absorbtion is still dissociative. While the
ground state Pt$_{5}$H$_{2}$ cluster (5p) and three other isomers
((5i), (5k), and (5m)) are in quintet states, the other six isomers
are in triplet states. Similar to the previous clusters, generally
bridge-top configurations have higher BEs than bridge-bridge
structures.

\subsection{\label{Frag}Fragmentation Behavior}

\begin{figure}
\includegraphics[bb= 0 0 36.76cm 42.51cm, scale=0.40]{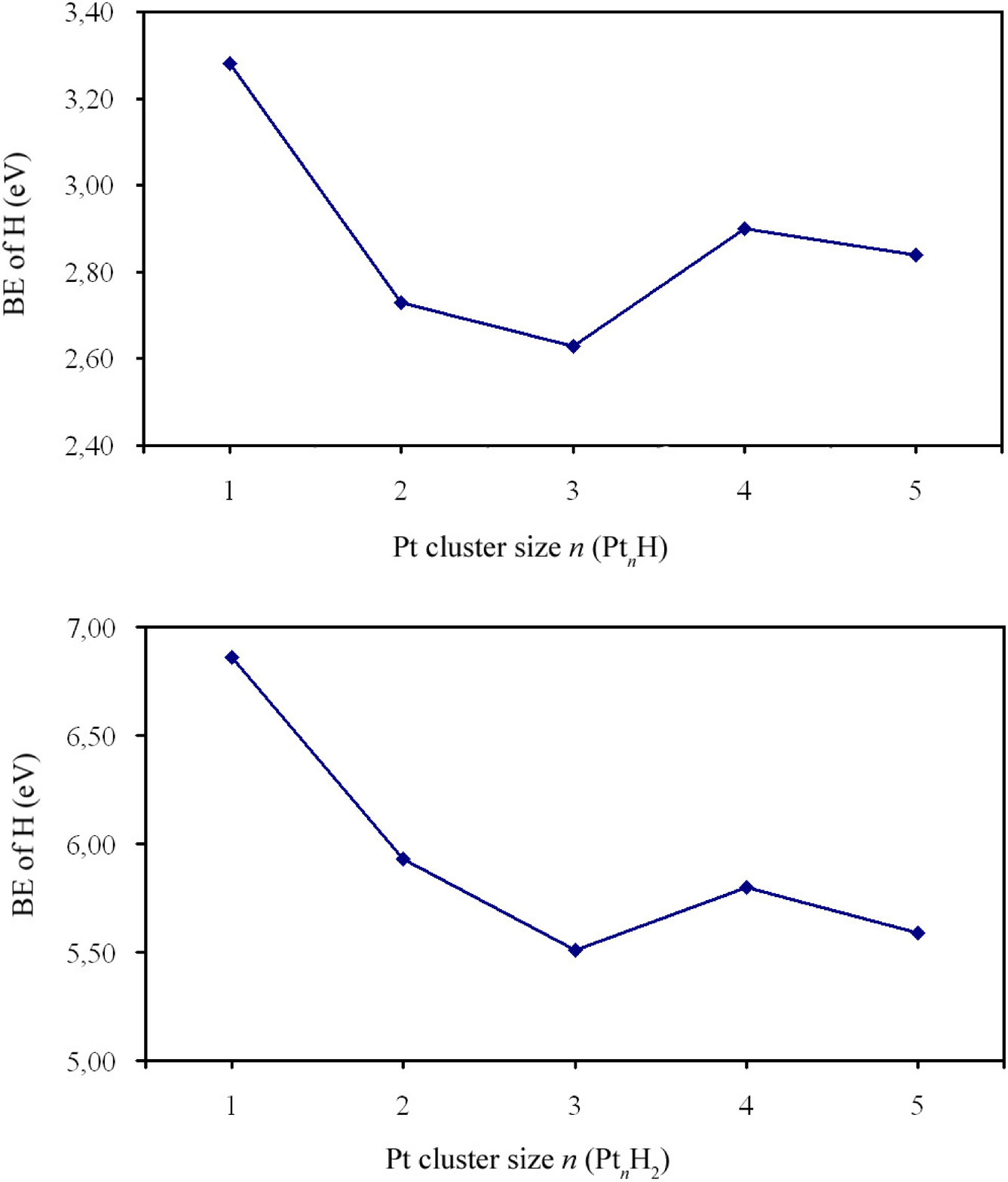}
\caption{\label{PtH} Binding energies of H (top) and 2H (bottom)
atoms on Pt$_{n}$ clusters.}
\end{figure}

Figure~\ref{PtH} shows the plot of the BE of one and two H atoms on
the lowest energy isomers of the Pt clusters. The H BE is large for
most clusters and in particular for clusters with 1 and 4 Pt atoms.
This is similar when 2H BEs are considered. The clusters with 1, and
4 Pt atoms have larger BEs for 2H compared to their neighboring
sizes. We have further studied the stability of these clusters from
the fragmentation channels in which a Pt atom, or a Pt$_{2}$, or PtH
molecule is one of the fragments. The fragmentation energies of all
of these channels are listed in Table~\ref{table4}. It is noted that
the fragmentation energies are the largest for Pt$_{4}$H$_{2}$ in Pt
and PtH fragmentation channels. Therefore, we expect it to be the
most stable species. For the Pt$_{2}$ fragmentation, Pt$_{5}$H$_{2}$
has the highest fragmentation energy. On the other hand,
Pt$_{2}$H$_{2}$ and Pt$_{3}$H$_{2}$ have the lowest fragmentation
energies for this channel.

\subsection{\label{Moment}Magnetic Moment of Pt$_{n}$H$_{m}$ Clusters}

In this section we discuss the change in the magnetic moment of the
hydrogenated platinum clusters when successive H atoms are added to
the bare Pt clusters. The magnetic moments and the majority and
minority spin LUMO energies of the lowest energy structures for
Pt$_{n}$H$_{m}$ ($n$ = 1-5, $m$ = 0-2) clusters are presented in
Table~\ref{table5}. It can be seen in Table~\ref{table5} that the
successive H addition to Pt atom and Pt dimer reduces the magnetic
moment as in the case of H absorption in bulk Pt. However, when the
successive H additions to the Pt trimer and pentamer are considered,
it is observed that as the magnetic moments decrease in the first
additions, they increase in the second ones. For the tetramer, the
magnetic moment increases in the first addition and decreases in the
second. Thus, the magnetic moments of certain small Pt clusters
exhibit an oscillatory change in the successive addition of H atoms.

The results obtained in this study confirm the explanation of Ashman
et al.~\cite{Khanna} that this situation can be understood by
looking at the LUMO energy differences between the majority and
minority spin electrons ($\delta E$ = minority spin LUMO energy -
majority spin LUMO energy). When the minority spin LUMO energy is
much less than the majority spin LUMO energy (when $\delta E$ is
highly negative), the additional electron goes to the spin state
with lowest LUMO, and reduces the magnetic moment to decrease the
cluster's total energy. This is what happens when successive H atoms
are added to Pt atom and dimer. However, if LUMO of minority is
slightly less than LUMO of majority (if $\delta E$ is greater than a
certain value which is around -0.227 eV for Pt-H clusters), the
additional electron may go to majority manifold since the exchange
coupling could lead to a rearrangement of the manifolds. This
explains the increases in the magnetic moments when H atoms are
added to the Pt$_{3}$H and Pt$_{4}$ clusters. It should be noted
here that nearly a half of the Pt$_{3}$H$_{2}$ isomers in
Table~\ref{table3} are in the singlet, and the others are in the
triplet states. Similarly, one of the Pt$_{4}$H isomer is in the
doublet and the other is in the quartet states. For Pt$_{5}$H, the
$\delta E$ is even positive and therefore we expect that H addition
will most probably increase the magnetic moment. In fact, this
expectation is confirmed in our results since the lowest energy
isomer of the Pt$_{5}$H$_{2}$ is found to be in the quintet state.

\section{SUMMARY}

In summary, we have presented results of studies on the interactions
in the bare and hydrogenated platinum clusters. We find that H
interacts strongly with Pt clusters. The interaction of H$_{2}$
molecule with any of the Pt clusters studied in the present work is
likely to be dissociative. H atoms can bond to Pt clusters either on
a bridge site or on a top site. We do not find any 3-fold bonding of
H to Pt clusters. In general, a top site bonding is more favorable
than a bridge site bonding. Therefore, for the bonding of two H
atoms, top-top configurations are more stable than bridge-top and
bridge-bridge configurations. Among all the clusters we have
investigated, we find that Pt$_{4}$H$_{2}$ is the best candidate to
be the most stable one. We do not agree with a previous claim that
3D geometries of Pt tetramer and pentamer are unfavored. We have
identified a distorted tetrahedron and a bridge site capped
tetrahedron as the lowest energy structures of Pt$_{4}$ and
Pt$_{5}$, respectively. We have also shown that the successive
addition of H atoms to Pt clusters decreases the magnetic moments of
the Pt atom and Pt dimer, whereas it leads to an oscillatory change
in the magnetic moment of Pt$_{3}$ - Pt$_{5}$ clusters.

\begin{acknowledgements}
This work is financially supported by The Scientific and
Technological Research Council of Turkey, Grant no. TBAG-HD/38
(105T051). I would also like to thank to Dr. Ersen Mete for
discussion and comments as well as his valuable computational
supports.
\end{acknowledgements}

\newpage

\begingroup
\squeezetable
\begin{table}
\caption{\label{table1}Comparison of the calculated and experiment
properties of H$_{2}$, Pt$_{2}$, and PtH. Experimental values for
the bond lengths, vibrational frequencies and binding energies of
H$_{2}$, Pt$_{2}$, and PtH are from Refs.~\cite{Huber1, Huber2},
Refs.~\cite{Airola,Fabbi,Taylor}, and
Refs.~\cite{Huber1,McCarthy}, respectively.}
\begin{ruledtabular}
\begin{tabular}{l c c c c c c c c c}
       & \multicolumn{3}{c}{Bond length ($\AA$)}  & \multicolumn{3}{c}{Vibrational frequency (cm$^{-1}$)}
       & \multicolumn{3}{c}{Binding energy (eV)} \\
\cline{2-4} \cline{5-7} \cline{8-10}
             & H$_{2}$ & Pt$_{2}$ & PtH  & H$_{2}$ & Pt$_{2}$ & PtH  & H$_{2}$ & Pt$_{2}$ & PtH  \\
\hline
Expt         & 0.74    & 2.33     & 1.53 & 4360    & 223      & 2294 & 4.75    & 3.14     & 3.44  \\
\hline
PW91PW91     & 0.75    & 2.37     & 1.53 & 4333    & 235      & 2423 & 4.27    & 3.51     & 3.35  \\
B3LYP        & 0.74    & 2.47     & 1.53 & 4419    & 205      & 2423 & 4.49    & 1.57     & 3.33  \\
SVWN5        & 0.77    & 2.33     & 1.52 & 4189    & 252      & 2475 & 4.64    & 4.37     & 4.12  \\
BPW91        & 0.75    & 2.37     & 1.53 & 4345    & 234      & 2420 & 4.30    & 3.36     & 3.28  \\
B98PW91      & 0.74    & 2.45     & 1.52 & 4439    & 210      & 2443 & 4.35    & 2.15     & 3.61  \\
\end{tabular}
\end{ruledtabular}
\end{table}
\endgroup

\begingroup
\squeezetable
\begin{table}
\caption{\label{table2}Isomeric structure properties of Pt$_n$
($n$=2-5) clusters.}
\begin{ruledtabular}
\begin{tabular}{l c c c c c c c}
     & Cluster      & Structure                      & Symmetry     & Ground         & Total BE& HOMO-LUMO &
     $\omega_l$ and $\omega_h$\footnote{Lowest and highest vibrational frequencies}\\
     &              &                                &              & state          & (eV)    & gap (eV)  & (cm$^{-1}$) \\
\hline
      & Pt$_{2}$    & linear                         &$D_{\infty h}$&$^{3}\Delta_{g}$& 3.36    & 1.32      & 234     \\
(3-1) & Pt$_{3}$    & equilateral triangle           & $D_{3h}$     & $^{3}E''$      & 6.57    & 1.05      & 118, 221   \\
(3-2) &             & isosceles triangle             & $C_{2v}$     & $^{3}B_{2}$    & 6.57    & 1.05      & 118, 220   \\
(4-1) & Pt$_{4}$    & distorted tetrahedron          & $C_{2}$      & $^{3}B$        & 9.76    & 0.44      & 74, 215    \\
(4-2) &             & out of plane rhombus           & $C_{2v}$     & $^{3}A_{2}$    & 9.66    & 0.40      & 28, 207    \\
(4-3) &             & Y-like                         & $C_{s}$      & $^{3}A''$      & 9.37    & 0.48      & 25, 262    \\
(5-1) & Pt$_{5}$    & bridge site capped tetrahedron & $C_{2}$      & $^{5}B$        & 13.20   & 0.69      & 29, 206    \\
(5-2) &             & rhombus pyramid                & $C_{2v}$     & $^{5}A_{2}$    & 13.11   & 0.30      & 32, 207    \\
(5-3) &             & trigonal bipyramid             & $C_{s}$      & $^{3}A''$      & 13.04   & 0.41      & 44, 207    \\
(5-4) &             & X-like                         & $D_{2h}$     & $^{3}B_{2u}$   & 13.00   & 0.60      & 22, 254    \\
(5-5) &             & W-like                         & $C_{1}$      & $^{3}A$        & 12.86   & 0.30      & 22, 227    \\
\end{tabular}
\end{ruledtabular}
\end{table}
\endgroup

\begingroup
\squeezetable
\begin{table}
\caption{\label{table3}Properties of Pt$_n$H$_m$ ($n$=1-5 and
$m$=0-2) clusters. Location of H is represented by symbols b and t
which mean bridge and top site, respectively.}
\begin{ruledtabular}
\begin{tabular}{l c c c c c c c c }
     & Cluster         & Location & Symmetry     & Ground          & Total BE& BE of H   & HOMO-LUMO &
     $\omega_l$ and $\omega_h$\footnote{Lowest and highest vibrational frequencies}\\
     &                 & of H     &              & state           & (eV)    & (eV)      & gap (eV)  & (cm$^{-1}$) \\
\hline
(1a) & PtH             &          &$C_{\infty v}$& $^{2}\Delta$    & 3.28    & 3.28      & 3.02      & 2420      \\
(1b) & PtH$_{2}$       &          & $C_{2v}$     & $^{1}A_{1}$     & 6.86    & 6.86      & 2.49      & 793, 2490 \\
(2a) & Pt$_{2}$        &          &$D_{\infty h}$& $^{3}\Delta_{g}$& 3.36    &           & 1.32      & 234       \\
(2b) & Pt$_{2}$H       & b        & $C_{2v}$     & $^{2}A_{2}$     & 6.09    & 2.73      & 1.26      & 192, 1636 \\
(2c) &                 & t        & $C_{s}$      & $^{4}A''$       & 5.61    & 2.25      & 2.57      & 183, 2213 \\
(2d) & Pt$_{2}$H$_{2}$ & b-b      & $C_{2v}$     & $^{1}A_{1}$     & 8.95    & 5.59      & 0.94      & 180, 1646 \\
(2e) &                 & t-t (1)  & $C_{2}$      & $^{1}A$         & 9.29    & 5.93      & 0.87      & 193, 2451 \\
(2f) &                 & t-t (2)  & $C_{s}$      & $^{1}A'$        & 8.49    & 5.13      & 0.09      & 202, 2397 \\
(3a) & Pt$_{3}$        &          & $D_{3h}$     & $^{3}E''$       & 6.57    &           & 1.05      & 118, 221  \\
(3b) & Pt$_{3}$H       & b        & $C_{s}$      & $^{2}A''$       & 9.20    & 2.63      & 0.66      & 115, 1715 \\
(3c) &                 & t        & $C_{1}$      & $^{2}A$         & 9.14    & 2.57      & 0.19      & 138, 2449 \\
(3d) & Pt$_{3}$H$_{2}$ & b-b (1)  & $C_{2v}$     & $^{1}A_{1}$     & 11.70   & 5.13      & 0.59      & 120, 1505 \\
(3e) &                 & b-b (2)  & $C_{2}$      & $^{3}A$         & 11.39   & 4.82      & 0.24      & 116, 1627 \\
(3f) &                 & b-b (3)  & $C_{s}$      & $^{1}A'$        & 11.80   & 5.23      & 0.50      & 62, 1499  \\
(3g) &                 & b-t (1)  & $C_{s}$      & $^{3}A''$       & 11.97   & 5.40      & 0.71      & 105, 2266 \\
(3h) &                 & b-t (2)  & $C_{1}$      & $^{1}A$         & 11.98   & 5.41      & 0.37      & 105, 2246 \\
(3i) &                 & b-t (3)  & $C_{s}$      & $^{1}A'$        & 11.57   & 5.00      & 0.17      & 113, 2237 \\
(3j) &                 & b-t (4)  & $C_{s}$      & $^{3}A''$       & 11.92   & 5.35      & 0.65      & 101, 2291 \\
(3k) &                 & t-t (1)  & $C_{2}$      & $^{1}A$         & 12.07   & 5.50      & 0.45      & 89, 2317  \\
(3l) &                 & t-t (2)  & $C_{s}$      & $^{3}A''$       & 12.08   & 5.51      & 1.03      & 107, 2316 \\
(4a) & Pt$_{4}$        &          & $C_{2}$      & $^{3}B$         & 9.76    &           & 0.44      & 74, 215   \\
(4b) & Pt$_{4}$H       & b        & $C_{1}$      & $^{2}A$         & 12.36   & 2.60      & 0.27      & 39, 1277  \\
(4c) &                 & t        & $C_{1}$      & $^{4}A$         & 12.66   & 2.90      & 0.59      & 83, 2232  \\
(4d) & Pt$_{4}$H$_{2}$ & b-b (1)  & $C_{s}$      & $^{3}A''$       & 15.05   & 5.29      & 0.38      & 68, 1334  \\
(4e) &                 & b-b (2)  & $C_{s}$      & $^{1}A'$        & 15.06   & 5.30      & 0.40      & 69, 1286  \\
(4f) &                 & b-t (1)  & $C_{1}$      & $^{3}A$         & 15.38   & 5.62      & 0.63      & 70, 2174  \\
(4g) &                 & b-t (2)  & $C_{1}$      & $^{3}A$         & 15.28   & 5.52      & 0.56      & 51, 2168  \\
(4h) &                 & t-t      & $C_{s}$      & $^{3}A'$        & 15.56   & 5.80      & 0.94      & 71, 2264  \\
(5a) & Pt$_{5}$        &          & $C_{2}$      & $^{5}B$         & 13.20   &           & 0.69      & 29, 206   \\
(5b) & Pt$_{5}$H       & b (1)    & $C_{2}$      & $^{4}B$         & 15.83   & 2.78      & 0.67      & 25, 1359  \\
(5c) &                 & b (2)    & $C_{1}$      & $^{4}A$         & 15.75   & 2.70      & 0.58      & 28, 1267  \\
(5d) &                 & b (3)    & $C_{1}$      & $^{4}A$         & 15.71   & 2.66      & 0.75      & 16, 1350  \\
(5e) &                 & t (1)    & $C_{1}$      & $^{4}A$         & 15.85   & 2.80      & 0.45      & 24, 2234  \\
(5f) &                 & t (2)    & $C_{1}$      & $^{4}A$         & 15.89   & 2.84      & 0.45      & 27, 2203  \\
(5g) & Pt$_{5}$H$_{2}$ & b-b (1)  & $C_{1}$      & $^{3}A$         & 18.32   & 5.27      & 0.49      & 27, 1397  \\
(5h) &                 & b-b (2)  & $C_{1}$      & $^{3}A$         & 18.29   & 5.09      & 0.67      & 17, 1364  \\
(5i) &                 & b-b (3)  & $C_{1}$      & $^{5}A$         & 18.44   & 5.24      & 0.47      & 43, 1344  \\
(5j) &                 & b-t (1)  & $C_{1}$      & $^{3}A$         & 18.49   & 5.29      & 0.41      & 26, 2232  \\
(5k) &                 & b-t (2)  & $C_{s}$      & $^{5}A''$       & 18.54   & 5.34      & 0.66      & 27, 2279  \\
(5l) &                 & b-t (3)  & $C_{1}$      & $^{3}A$         & 18.54   & 5.34      & 0.43      & 27, 2207  \\
(5m) &                 & b-t (4)  & $C_{1}$      & $^{5}A$         & 18.31   & 5.26      & 0.52      & 19, 2263  \\
(5n) &                 & t-t (1)  & $C_{1}$      & $^{3}A$         & 18.55   & 5.35      & 0.32      & 17, 2261  \\
(5o) &                 & t-t (2)  & $C_{1}$      & $^{3}A$         & 18.64   & 5.44      & 0.29      & 29, 2288  \\
(5p) &                 & t-t (3)  & $C_{s}$      & $^{5}A''$       & 18.79   & 5.59      & 0.67      & 64, 2323  \\
\end{tabular}
\end{ruledtabular}
\end{table}
\endgroup

\begingroup
\squeezetable
\begin{table}
\caption{\label{table4}Fragmentation energies of Pt$_{n}$H$_{m}$
clusters. Energies are given in eV and a positive value means that
the parent cluster has a larger BE than the sum of the BEs of the
products.}
\begin{ruledtabular}
\begin{tabular}{l c c c c }
       & Cluster (Pt$_{n}$H$_{m}$)  & Pt$_{n-1}$H$_{m}$ + Pt & Pt$_{n-2}$H$_{m}$ + Pt$_{2}$ & Pt$_{n-1}$H$_{m-1}$ + PtH \\
\hline
(2b)   & Pt$_{2}$H                  & 2.81                   & 2.73                         &  2.81 \\
(2e)   & Pt$_{2}$H$_{2}$            & 2.43                   & 1.63                         &  2.73 \\
(3b)   & Pt$_{3}$H                  & 3.12                   & 2.57                         &  2.57 \\
(3l)   & Pt$_{3}$H$_{2}$            & 2.79                   & 1.86                         &  2.71 \\
(4c)   & Pt$_{4}$H                  & 3.45                   & 3.21                         &  2.81 \\
(4h)   & Pt$_{4}$H$_{2}$            & 3.48                   & 2.91                         &  3.07 \\
(5f)   & Pt$_{5}$H                  & 3.23                   & 3.32                         &  2.85 \\
(5p)   & Pt$_{5}$H$_{2}$            & 3.23                   & 3.35                         &  2.85 \\
\end{tabular}
\end{ruledtabular}
\end{table}
\endgroup

\begingroup
\squeezetable
\begin{table}
\caption{\label{table5}The majority and minority LUMO levels}
\begin{ruledtabular}
\begin{tabular}{l c c c c c}
       & Cluster         & Magnetic     & Majority & Minority & $\delta E$ \\
       &                 & moment       & LUMO (eV)& LUMO (eV)& (eV)       \\
\hline
       & Pt              &   2          & -1.155   & -5.430   & -4.275 \\
(1a)   & PtH             &   1          & -2.660   & -5.150   & -2.490 \\
(1b)   & PtH$_{2}$       &   0          & -2.821   & -2.821   &  0.000 \\
(2a)   & Pt$_{2}$        &   2          & -4.427   & -5.108   & -0.681 \\
(2b)   & Pt$_{2}$H       &   1          & -4.127   & -5.059   & -0.932 \\
(2e)   & Pt$_{2}$H$_{2}$ &   0          & -4.426   & -4.426   &  0.000 \\
(3a)   & Pt$_{3}$        &   2          & -4.083   & -4.803   & -0.720 \\
(3b)   & Pt$_{3}$H       &   1          & -4.201   & -4.428   & -0.227 \\
(3l)   & Pt$_{3}$H$_{2}$ &   2          & -4.385   & -5.124   & -0.739 \\
(4a)   & Pt$_{4}$        &   2          & -4.104   & -4.215   & -0.111 \\
(4c)   & Pt$_{4}$H       &   3          & -4.401   & -4.552   & -0.151 \\
(4h)   & Pt$_{4}$H$_{2}$ &   2          & -4.345   & -4.524   & -0.179 \\
(5a)   & Pt$_{5}$        &   4          & -4.428   & -4.600   & -0.172 \\
(5f)   & Pt$_{5}$H       &   3          & -4.743   & -4.682   &  0.061 \\
(5p)   & Pt$_{5}$H$_{2}$ &   4          & -4.699   & -5.087   & -0.388 \\
\end{tabular}
\end{ruledtabular}
\end{table}
\endgroup

\newpage

\begin{thebibliography}{abdx}

\bibitem{Okamoto}Y. Okamoto, Chem. Phys. Lett. 405, 79 (2005).

\bibitem{Poulain}E. Poulain, J. Benitez, S. Castillo, V. Bertin, A. Cruz,
J. Mol. Struc. THEOCHEM 709, 67 (2004).

\bibitem{Bartczak}W.M. Bartczak, J. Stawowska, Struc. Chem. 15, 447
(2004).

\bibitem{Khanna}C. Ashman, S.N. Khanna, M.R. Pederson, Chem. Phys.
Lett. 368, 257 (2003).

\bibitem{Musaev}J. Moc, D.G. Musaev, K.J. Morokuma, J. Phys. Chem. A 107, 4929
(2003).

\bibitem{Mainardi}D.S. Mainardi, P.B. Balbuena, J. Phys. Chem. A. 107, 10370
(2003).

\bibitem{Kumar}H. Kawamura, V. Kumar, Q. Sun, Y. Kawazoe, Phys.
Rev. B 65, 045406 (2001).

\bibitem{Ali}A. Sebetci and Z.B. G\"{u}ven\c{c}, Surf. Sci.
525, 66 (2003); Eur. Phys. J. D, 30, 71 (2004);  Modelling Simul.
Mater. Sci. Eng. 13, 683, (2005).

\bibitem{Gronbeck}H. Gr\"{o}nbeck, W. Andreoni, Chem. Phys. 262, 1 (2000).

\bibitem{Xiao}L. Xiao, L. Wang, J. Phys. Chem. A 108, 8605 (2004).

\bibitem{Tian}W.Q. Tian, M. Ge, B.R. Sahu, D. Wang, T. Yamada, S. Mashiko,
J. Phys. Chem. A 108, 3806 (2004).

\bibitem{Poulain2}E. Poulain, J. Garcia-Prieto, M.E. Ruiz, O. Novaro,
Int. J. Quantum. Chem. 29, 1181 (1986).

\bibitem{Poulain3}E. Poulain, V. Bertin, S. Castillo, A. Cruz,
J. Mol. Catal. A: Chem. 116, 385 (1997).

\bibitem{Nakasuji}H. Nakasuji, Y. Matsuzaki, T. Yonezawa, J. Chem.
Phys. 88, 5759 (1988).

\bibitem{Bala1}K. Balasubramanian, J. Chem.
Phys. 87, 2800 (1987);  J. Chem. Pyhs. 87, 6573 (1987); J. Chem.
Pyhs. 94, 1253 (1991).

\bibitem{Bala2}K. Balasubramanian, P. Y. Feng, J. Chem.
Phys. 92, 541 (1990).

\bibitem{Dai}D. Dai, W. Liao, K. Balasubramanian, J. Chem. Phys.
102, 7530 (1995).

\bibitem{Cui}Q. Cui, D.G. Musaev, K. Morokuma, J. Phys. Chem. A
108, 8418 (1998).

\bibitem{Samson}P. Samson, A. Nesbitt, B.E. Koel, A. Hodgson, J. Chem. Phys
109, 3255 (1998).

\bibitem{Kaldor}A. Kaldor, D.M. Cox, M. R. Zakin, Adv. Chem. Phys.
70, 211 (1988).

\bibitem{Andrews}L. Andrews, X. Wang, L. Manceron, J. Chem. Phys.
114, 1559 (2001).

\bibitem{Knick}M.B. Knickelbein, G.M. Koretsky, K.A. Jackson, M.R.
Pederson, Z. Hajnal, J. Chem. Phys. 109, 10692 (1998).

\bibitem{Luntz}A.C. Luntz, J.K. Brown, M.D. Williams, J. Chem. 93,
5240 (1990).

\bibitem{NWChem} E. Apra, T. L. Windus, T. P. Straatsma et al.
NWChem, A Computational Chemistry Package for Parallel Computers,
Version 4.7 (2005).
% E. J. Bylaska, W. de Jong,
% S. Hirata, M. Valiev, M. T. Hackler, L. Pollack, K. Kowalski,
% R. J. Harrison, M. Dupuis, D. M. A. Smith, J. Nieplocha,
% V. Tipparaju, M. Krishnan, A. A. Auer, E. Brown, G. Cisneros,
% G. I. Fann, H. Fruchtl, J. Garza, K. Hirao, R. Kendall,
% J. A. Nichols, K. Tsemekhman, K. Wolinski, J. Anchell, D. Bernholdt,
% P. Borowski, T. Clark, D. Clerc, H. Dachsel, M. Deegan, K. Dyall,
% D. Elwood, E. Glendening, M. Gutowski, A. Hess, J. Jaffe, B. Johnson,
% J. Ju, R. Kobayashi, R. Kutteh, Z. Lin, R. Littlefield, X. Long,
% B. Meng, T. Nakajima, S. Niu, M. Rosing, G. Sandrone, M. Stave,
% H. Taylor, G. Thomas, J. van Lenthe, A. Wong, and Z. Zhang,

\bibitem{LAN}P.J. Hay, W.R. Wadt, J. Chem. Phys. 82, 299 (1985).

\bibitem{Perdew}J.P. Perdew, J.A. Chevary, S.H. Vosko, K.A.
Jackson, M.R. Pederson, D.J. Singh, C. Fiolhais, Phys. Rev. B 46,
6671 (1992).

\bibitem{Becke1}A.D. Becke, Phys. Rev. A 38, 3098 (1988).

\bibitem{Slater}J.C. Slater, Quantum Theory of Molecules and
Solids (McGraw-Hill, New York, 1974).

\bibitem{Becke2}A.D. Becke, J. Chem. Phys. 98, 5648 (1993).

\bibitem{Becke3}H.L. Schmider, A.D. Becke, J. Chem. Phys. 108, 9624 (1998).

\bibitem{ChemCraft}http://www.chemcraftprog.com.

\bibitem{Huber1}K.P. Huber, G. Herzberg, Molecular Spectra and
Molecular Structure. IV. Constants of Diatomic Molecules (Van
Nostrand Reinhold, New York, 1979).

\bibitem{Huber2}K.P. Huber, in American Institute of Physics Handbook,
edited by D.E. Gray (McGraw-Hill, New York, 1972).

\bibitem{Airola}M.B. Airola, M.D. Morse, J. Chem. Phys. 116, 1313
(2002).

\bibitem{Fabbi}J.C. Fabbi, J.D. Langenberg, Q.D. Costello, M.D.
Morse, L. Karlsson, J. Chem. Phys. 115, 7543 (2001).

\bibitem{Taylor}S. Taylor, G.W. Lemire, Y.M. Hamrick, Z. Fu,
M.D. Morse, J. Chem. Phys. 89, 5517 (1988).

\bibitem{McCarthy}M.C. McCarthy, R.W. Field, R. Engleman, P.F.
Bernath, J. Mol. Spectrosc. 158, 208 (1993).

\bibitem{Dai2}D. Dai, K. Balasubramanian, J. Chem. Phys.
103, 648 (1995).

%\bibitem{Moore}C.E. Moore , Atomic Energy Levels, National Bureau of
%Standard 467, (1958).

%\bibitem{Harms}O. Harms, M. Zehnpfennig, V. Gomer, D. Meschede,
%J. Phys. B 30, 3781 (1997).

%\bibitem{Marijnissen}A. Marijnissen, J.J. Termeulen, P.A. Hackett, B. Simard,
%Phys. Rev. A 52, 2606 (1995).

%\bibitem{Bilodeau}R.C. Bilodeau, M. Scheer, H.K. Haugen, R.L.
%Brooks, Phys. Rev. A 61, 012505 (2000).

%\bibitem{Janak}J.F. Janak, Phys. Rev. B 18, 7165 (1978).

%\bibitem{Casida}M.E. Casida, C. Jamorski, K.C. Casida, D.R.
%Salahub, J. Chem. Phys. 108, 4439 (1998).

\end{thebibliography}
\end{document}